\documentclass[superscriptaddress,preprint]{revtex4}

\usepackage[tbtags]{amsmath}
\usepackage{amsfonts}
\usepackage{mathrsfs}
\usepackage{amssymb}
\usepackage{graphicx}
\usepackage{epsfig,graphicx,times}
\usepackage{subfigure}
\usepackage{color}
\usepackage{amssymb}
\usepackage{amsmath}
\usepackage{graphicx}
\usepackage{dcolumn}
\usepackage{bm}
\usepackage{float}
\usepackage[abs]{overpic}

\begin{document}

\title{Dissipative dynamics of few-photons superposition states:
A dynamical invariant}
\author{Hong-Yan Wen}
\affiliation{Quantum Optoelectronics Laboratory, School of Physics
and Technology, Southwest Jiaotong University, Chengdu 610031,
China}
\author{Jing Cheng}
\affiliation{Department of Physics, South China University of
Technology, Guangzhou 510640, China}
\author{Y. Yang}
\affiliation{State Key Laboratory of Optoelectronic Materials and
Technologies, School of Physics and Engineering, Sun Yat-Sen
University, Guangzhou 510275, China}
\author{L. F. Wei\footnote{weilianfu@gmail.com; weilianf@mail.sysu.edu.cn}}
\affiliation{Quantum Optoelectronics Laboratory, School of Physics
and Technology, Southwest Jiaotong University, Chengdu 610031,
China} \affiliation{State Key Laboratory of Optoelectronic Materials
and Technologies, School of Physics and Engineering, Sun Yat-Sen
University, Guangzhou 510275, China}
\date{\today}

\begin{abstract}
By numerically calculating the time-evolved Wigner functions, we
investigate the dynamics of a few-photon superposed (e.g., up to two
ones) state in a dissipating cavity. It is shown that, the
negativity of the Wigner function of the photonic state
unquestionably vanishes with the cavity's dissipation. As a
consequence, the nonclassical effects related to the negativity of
the Wigner function should be weakened gradually. However, it is
found that the value of the second-order correlation function
$g^{(2)}(0)$ (which serves usually as the standard criterion of a
typical nonclassical effect, i.e., $g^{(2)}(0)<1$ implies that the
photon is anti-bunching) is a dynamical invariant during the
dissipative process of the cavity. This feature is also proven
analytically and suggests that $g^{(2)}(0)$ might not be a good
physical parameter to describe the photonic decays. Alternatively,
we find that the anti-normal-order correlation function
$g^{(2A)}(0)$ changes with the cavity's dissipation and thus is more
suitable to describe the dissipative-dependent cavity. Finally, we
propose an experimental approach to test the above arguments with a
practically-existing cavity QED system.

\vspace{0.2cm}

\end{abstract}

\maketitle

\section{Introduction}

It is well-known that the Wigner function, introduced 70 years ago
by Wigner to describe the quasi-probability distribution of a
quantum particle in its phase space, is a very popular tool to study
the statistical properties of various quantum states [1]. Basically,
once the Wigner function has been determined, all the knowable
information on the quantum state (such as its nonclassical
statistical properties) can be extracted [2-5]. Typically, differing
from the standard probability distribution, such a quasi-probability
distribution can be assigned by a negative value [6]. Therefore, a
quantum state with the negative Wigner function should be
nonclassical and thus certain nonclassical effects (such as the
photon anti-buchings) [7-11] demonstrate. This indicates that,
determining the Wigner function of a selected quantum state plays an
important role both fundamentally and practically in quantum-state
engineerings.

Usually, any selected quantum system is always surrounded by the
classical environments. Thus, dissipation of the
artificially-prepared quantum sate is one of the central topics in
quantum coherence science. Roughly, due to the existence of various
dissipations and fluctuations from the environments, any excited
quantum state will decay to the ground state and the relevant system
finally becomes classical. Under the standard logic, people pay the
most attention to calculate either decoherence or decay time of a
superposition quantum state, rather than cares on the process of the
decoherence or decay [12-15]. Alternatively, in the present work we
investigate exactly the dissipative dynamics for a prepared quantum
state by calculating its dissipative-dependent Wigner function. Our
discussions are based on the typical few-photon quantum state in a
cavity, but can be directly generalized to other quantum systems
such as qubits and qutrits.

The paper is organized as: in Sec.~2, we describe how the Wigner
function for a few-photon superposed state changes with the cavity's
dissipation. Our numerical results show naturally that the
negativity of the Wigner function weakens gradually with the
dissipation and the final state of the cavity should be ``classical"
with {\it positive} Wigner function.
With the calculated Wigner function we investigate how the
nonclassical properties, such as the anti-bunching effect of
photons, changes with the cavity dissipation. It is surprised that
the value of the second-order correlation function $g^{(2)}(0)$
(which serves usually as the standard criterion of a nonclassical
effect, i.e., $g^{(2)}(0)<1$ implies that the photon is
anti-bunching) is an invariant during the dissipative process of the
cavity. We prove such an argument analytically by directly solving
the relevant master equation and suggests that $g^{(2)}(0)$ is not a
good parameter to describe dissipative-dependent non-classicality of
the photonic decays. Alternatively, we find that the
anti-normal-order correlation function $g^{(2A)}(0)$ changes with
the cavity's dissipation and thus could be more suitable to describe
the dissipative-dependent cavity.
With an experimentally-demonstrated cavity QED system we propose an
approach to test our results, including how to prepare the
investigated few-photon superposed state of the cavity and measure
its Wigner function. Finally, our conclusions and discussions are
given in Sec.~4.

\section{Dissipative dynamics of Wigner functions for few-photons superposition states}

Generally, the quasi-probability distribution $W(\alpha,\alpha^{*})$
can be defined by the Fourier transform of the symmetrical-ordered
characteristic function $C(\lambda,\lambda^{*})$ [16], i.e.,
 \begin{equation}
W(\alpha,\alpha^{*})=\frac{1}{\pi^{2}}\int d^{2}\lambda
\,\,C(\lambda,\lambda^{*})e^{\alpha\lambda^{*}-\alpha^{*}\lambda},
\end{equation}
with $\lambda$ and $\alpha$ being the complex parameters in phase
space. The expression of the symmetrical-ordered characteristic
function is defined as
\begin{equation}
C(\lambda,\lambda^{*})=Tr[\rho
e^{\lambda\hat{a}^\dagger-\hat{a}\lambda^{*}}],
\end{equation}
where $\rho$ is the density matrix of the cavity state
$|\psi\rangle$, and $\hat{a}$ and $\hat{a}^\dagger$ the usual
annihilation and creation operators of the photons, respectively.

For the simplicity and without loss of the generality, let us assume
that the cavity is initially prepared in the following few-photon
superposition state
\begin{equation}
|\psi(0)\rangle=C_{0}|0\rangle+C_{1}|1\rangle+C_{2}|2\rangle,
\end{equation}
with the complex amplitudes: $C_{0}=|C_{0}|e^{i\phi},
C_{1}=|C_{1}|$, and $C_{2}=|C_{2}|e^{i\varphi}$.
Then, with the matrix elements of Wigner operator:
$\vartriangle(\alpha,\alpha^{*}) =\int
d^{2}ze^{[z(\hat{a}^{\dagger}-\alpha^{*})
-z^{*}(\hat{a}-\alpha)]}/2\pi^{2}$, in the Fock representation [17]
\begin{equation}
\langle n|\vartriangle(\alpha,\alpha^{*})|m\rangle
=\frac{(-1)^{m}}{\pi}\sqrt{\frac{m!}{n!}}(2\alpha)^{n-m}
e^{(-2|\alpha|^{2})}L^{(n-m)}_{m}(4|\alpha|^{2}),\,n,m=0,1,2,...
\end{equation}
here $n>m, \alpha_{0}=|\alpha_{0}|e^{i\theta}$, one can easily
obtain the Wigner function of the initial state
\begin{eqnarray}
W(\alpha_{0},\alpha^{*}_{0},0)&=&\frac{2}{\pi}[|C_{0}|^{2}-|C_{1}|^{2}L^{0}_{1}(4|\alpha_{0}|^{2})
+|C_{2}|^{2}L^{0}_{2}(4|\alpha_{0}|^{2})]e^{(-2|\alpha_{0}|^{2})}\nonumber\\
&+&\frac{8\sqrt{2}}{\pi}e^{(-2|\alpha_{0}|^{2})}|C_{0}C_{2}||\alpha_{0}|^{2}
\cos(2\theta-\varphi+\phi)\nonumber\\
&-&\frac{4\sqrt{2}}{\pi}e^{(-2|\alpha_{0}|^{2})}|C_{1}C_{2}||\alpha_{0}|
\cos(\theta-\varphi)L^{1}_{1}(4|\alpha_{0}|^{2})\nonumber\\
&+&\frac{8}{\pi}e^{(-2|\alpha_{0}|^{2})}|C_{0}C_{1}||\alpha_{0}|\cos(\theta+\phi).
\end{eqnarray}
for the above superposition initial state. Above, $L^{J}_{n}(x)$ is
an associated Laguerre polynomial defined by [18]
\begin{equation}
L^{(J)}_{n}(x)=\sum\limits_{\kappa=0}^n(-1)^{\kappa}\frac{(n+J)!}{(n-\kappa)!(J+\kappa)!}\frac{x^{\kappa}}{\kappa!}.
\end{equation}

In what follows we discuss how such a state decay in a loss cavity
by investigating the time-evolutions of the above initial Wigner
function.

\subsection{Dissipative dynamics for the Wigner function}

We now consider how the above few-photons superposition state
dissipates in a loss cavity without any thermal photon (i.e.,
$\langle n\rangle_{\rm th}=1/[\exp(\hbar\nu/k_BT)-1]\rightarrow 0$,
for the present optical frequency photons and at the room
temperature: $\hbar\nu/k_BT\gg 1$), which is described simply by the
following master equation [19-20]
\begin{equation}
\frac{d\rho}{dt}=-\kappa(\hat{a}^{\dagger}\hat{a}\rho+\rho\hat{a}^{\dagger}\hat{a}-2\hat{a}\rho\hat{a}^{\dagger}),
\end{equation}
with $k$ being the loss coefficient.
Our discussion is based on the time-evolutions of the Wigner
function, i.e.,
\begin{equation}
\frac{d}{dt}W(\alpha,\alpha^{*})=\frac{1}{\pi^{2}}\int
d^{2}\lambda\,\,\frac{d\,C(\lambda,\lambda^{*})}{dt}e^{\alpha\lambda^{*}
-\alpha^{*}\lambda},
\end{equation}
with
\begin{equation}
\frac{d\,C(\lambda,\lambda^{*})}{dt}=Tr[\frac{d\rho}{dt}
e^{\lambda\hat{a}^{\dagger}-\hat{a}\lambda^{*}}]=\kappa
Tr[(2\hat{a}\rho\hat{a}^{\dagger}
-\hat{a}^{\dagger}\hat{a}\rho-\rho\hat{a}^{\dagger}\hat{a})
e^{\lambda\hat{a}^\dagger-\hat{a}\lambda^{*}}].
\end{equation}
Formally, Eq.~(8) can be rewritten as
\begin{eqnarray}
\frac{d}{dt}W(\alpha,\alpha^{*}) =2\kappa
W^{[\hat{a}\rho\hat{a}^{\dagger}]}(\alpha,\alpha^{*}) -\kappa
W^{[\hat{a}^{\dagger}\hat{a}\rho]}(\alpha,\alpha^{*}) -\kappa
W^{[\rho\hat{a}^{\dagger}\hat{a}]}(\alpha,\alpha^{*}),
\end{eqnarray}
where the symbol $W^{[x]}(\alpha,\alpha^{*})$ is defined as
\begin{eqnarray}
W^{[x]}(\alpha,\alpha^{*})&=&\frac{1}{\pi^{2}}\int d^{2}\lambda
\,\,C^{[x]}(\lambda,\lambda^{*},t)e^{\alpha\lambda^{*}-\alpha^{*}\lambda},\,\,
C^{[x]}(\lambda,\lambda^{*})=Tr[x\,e^{\lambda\hat{a}^\dagger-\hat{a}\lambda^{*}}],
\end{eqnarray}
with $W^{[\rho]}(\alpha,\alpha^{*})=W(\alpha,\alpha^{*})$, and
$C^{[\rho]}(\lambda,\lambda^{*})=C(\lambda,\lambda^{*})$.
Note that
\begin{eqnarray}
C^{[\rho\hat{a}^{\dagger}\hat{a}]}(\lambda,\lambda^{*})&=&[\frac{1}{2}
+\frac{\partial}{\partial\lambda}(-\frac{\partial}{\partial\lambda^{*}})]C(\lambda,\lambda^{*})
=(\frac{\partial}{\partial\lambda}+\frac{\lambda^{*}}{2})
(\frac{\lambda}{2}-\frac{\partial}{\partial\lambda^{*}})C(\lambda,\lambda^{*}),
\end{eqnarray}
and
\begin{eqnarray}
\frac{1}{\pi^{2}}\int d^{2}\lambda
\,\,e^{(\alpha\lambda^{*}-\alpha^{*}\lambda)}
\frac{\partial}{\partial\lambda}C(\lambda,\lambda^{*})
&=&\alpha^{*}W(\alpha,\alpha^{*}),\nonumber\\
\frac{1}{\pi^{2}}\int d^{2}\lambda
\,\,e^{(\alpha\lambda^{*}-\alpha^{*}\lambda)}
\frac{\partial}{\partial\lambda^{*}}C(\lambda,\lambda^{*})
&=&-\alpha W(\alpha,\alpha^{*}),\nonumber\\
\frac{1}{\pi^{2}}\int d^{2}\lambda
\,\,e^{(\alpha\lambda^{*}-\alpha^{*}\lambda)}
\lambda^{*}C(\lambda,\lambda^{*})&=&\frac{\partial}{\partial\alpha}W(\alpha,\alpha^{*}),\nonumber\\
\frac{1}{\pi^{2}}\int d^{2}\lambda \,\,(-\lambda)
e^{(\alpha\lambda^{*}-\alpha^{*}\lambda)}
C(\lambda,\lambda^{*})&=&\frac{\partial}{\partial\alpha^{*}}W(\alpha,\alpha^{*}),
\end{eqnarray}
we then have
\begin{eqnarray}
W^{[\hat{a}\rho\hat{a}^{\dagger}]}(\alpha,\alpha^{*})&=&\frac{1}{\pi^{2}}\int
d^{2}\lambda\,\,C^{[\hat{a}\rho\hat{a}^{\dagger}]}(\lambda,\lambda^{*})e^{\alpha\lambda^{*}
-\alpha^{*}\lambda}\nonumber\\
&=&\frac{1}{\pi^{2}}\int
d^{2}\lambda\,\,[\alpha\alpha^{*}+\frac{1}{2}-
\alpha^{*}\frac{\partial}{\partial\alpha^{*}}-
\frac{1}{4}\frac{\partial}{\partial\alpha}\frac{\partial}{\partial\alpha^{*}}
+\frac{\alpha}{2}\frac{\partial}{\partial\alpha}]C(\lambda,\lambda^{*})e^{\alpha\lambda^{*}
-\alpha^{*}\lambda}\nonumber\\
&=&\frac{1}{\pi^{2}}\int
d^{2}\lambda\,\,[\alpha^{*}+\frac{1}{2}\frac{\partial}{\partial\alpha}]
[\alpha-\frac{1}{2}\frac{\partial}{\partial\alpha^{*}}]C(\lambda,\lambda^{*})e^{\alpha\lambda^{*}
-\alpha^{*}\lambda}\nonumber\\
&=&[\alpha^{*}+\frac{1}{2}\frac{\partial}{\partial\alpha}]
[\alpha-\frac{1}{2}\frac{\partial}{\partial\alpha^{*}}]W(\alpha,\alpha^{*}).
\end{eqnarray}
Similarly,
\begin{eqnarray}
W^{[\hat{a}\rho\hat{a}^{\dagger}]}(\alpha,\alpha^{*})&=&[\alpha+\frac{1}{2}\frac{\partial}{\partial\alpha^{*}}]
[\alpha^{*}+\frac{1}{2}\frac{\partial}{\partial\alpha}]W^{[\rho]}(\alpha,\alpha^{*})\nonumber\\
W^{[\hat{a}^{\dagger}\hat{a}\rho]}(\alpha,\alpha^{*})&=&[\alpha^{*}-\frac{1}{2}\frac{\partial}{\partial\alpha}]
[\alpha+\frac{1}{2}\frac{\partial}{\partial\alpha^{*}}]W^{[\rho]}(\alpha,\alpha^{*}).
\end{eqnarray}
As a consequence, Eq.~(10) reduces to
\begin{equation}
\frac{d W(\alpha,\alpha^{*})}{d t}=k[\frac{\partial^{2}}{\partial
\alpha\partial \alpha^{*}}+\frac{\partial}{\partial
\alpha}\alpha+\frac{\partial}{\partial
\alpha^{*}}\alpha^{*}]W(\alpha,\alpha^{*}),
\end{equation}
whose solution reads [12]
\begin{equation}
W(\alpha,\alpha^{*},t)=\frac{2}{1-e^{-2\kappa t}}\int
\frac{d^{2}\alpha_{0}}{\pi}e^{[-\frac{2}{1-e^{-2\kappa
t}}|\alpha-\alpha_{0}e^{-\kappa
t}|^{2}]}W(\alpha_{0},\alpha^{*}_{0},0),
\end{equation}

For the cavity initial state $|\psi(0)\rangle$ we substitute Eq.~(5)
into Eq.~(17) and get
\begin{eqnarray}
W(\alpha,\alpha^{*},t)&=&\frac{2}{\pi}e^{(-2|\alpha|^{2})}[|C_{0}|^{2}-|C_{1}|^{2}(2e^{-2\kappa
t}-1)]L_{1}^{0}[-\frac{|2\alpha
e^{-2\kappa t}|^{2}}{1-2e^{-2\kappa t}}]\nonumber\\
&+&\frac{2}{\pi}e^{-2|\alpha|^{2}}|C_{2}|^{2}(2e^{-2\kappa
t}-1)^{2}L_{2}^{0}[-\frac{|2\alpha
e^{-2\kappa t}|^{2}}{1-2e^{-2\kappa t}}]\nonumber\\
&+&\frac{8\sqrt{2}}{\pi}|C_{0}C_{2}|e^{(-2|\alpha|^{2}-2\kappa t)}|\alpha|^{2}\cos(2\theta-\varphi+\phi)\nonumber\\
&+&\frac{8}{\pi}|C_{0}C_{1}|e^{(-2|\alpha|^{2}-\kappa t)}|\alpha|\cos(\theta+\phi)\nonumber\\
&+&\frac{8\sqrt{2}}{\pi}|C_{1}C_{2}|e^{(-2|\alpha|^{2}-k
t)}|\alpha|\cos(\theta-\varphi)[2(|\alpha|^{2}-1)e^{-2\kappa t}+1].
\end{eqnarray}
Above, an integral formula [21]
\begin{eqnarray}
\int&&\frac{d^{2}z}{\pi}z^{n}z^{*m}e^{\{x_{1}|z|^{2}+x_{2}z+x_{3}z^{*}\}}\nonumber\\
&&=e^{(-\frac{x_{2}x_{3}}{x_{1}})}
\sum\limits_{\kappa=0}^{min(m,n)}\frac{n!m!}{\kappa!(n-\kappa)!
(m-\kappa)!(-x_{1})^{m+n-\kappa+1}}x_{2}^{m-\kappa}x_{3}^{n-\kappa},\,\,Re(x_{1})<0,
\end{eqnarray}
has been used and the unassociated Laguerre Polynomial $L_{m}(x,y)$:
\begin{eqnarray}
&&L_{m}(x,y)=\frac{(-1)^{m}}{m!}H_{m,n}(x,y),\,\,
H_{m,n}=\frac{\partial^{m+n}}{\partial T^{m}\partial
T'^{n}}e^{[-TT'+Tx+T'y]}|_{T=T'=0},
\end{eqnarray}
was introduced~[22-23] with $H_{m,n}(x,y)$ being the generating
function of two-variable Hermite polynomial.

\subsection{Time-dependent negativity of the Wigner function}
With the above time-evolution Wigner function, we next check how its
negativity changes with the cavity loss. Fig.~1 numerically shows
these changes with the effective time $\kappa t$ for the parameters:
$|C_{1}|=1/3, |C_{2}|=\sqrt{2}/2, \theta=\varphi=\pi, \phi=0$. Here,
for convenience we define the Wigner function $W(x,p,t)$ in the
$(x,p)$-space with $x=(\alpha+\alpha^*)/2$ and
$p=(\alpha-\alpha^*)/(2i)$. One can see:

\begin{figure*}[t]
\begin{minipage}{1.0\textwidth}
\centering
\includegraphics[width=16.0cm,height=12.0cm]{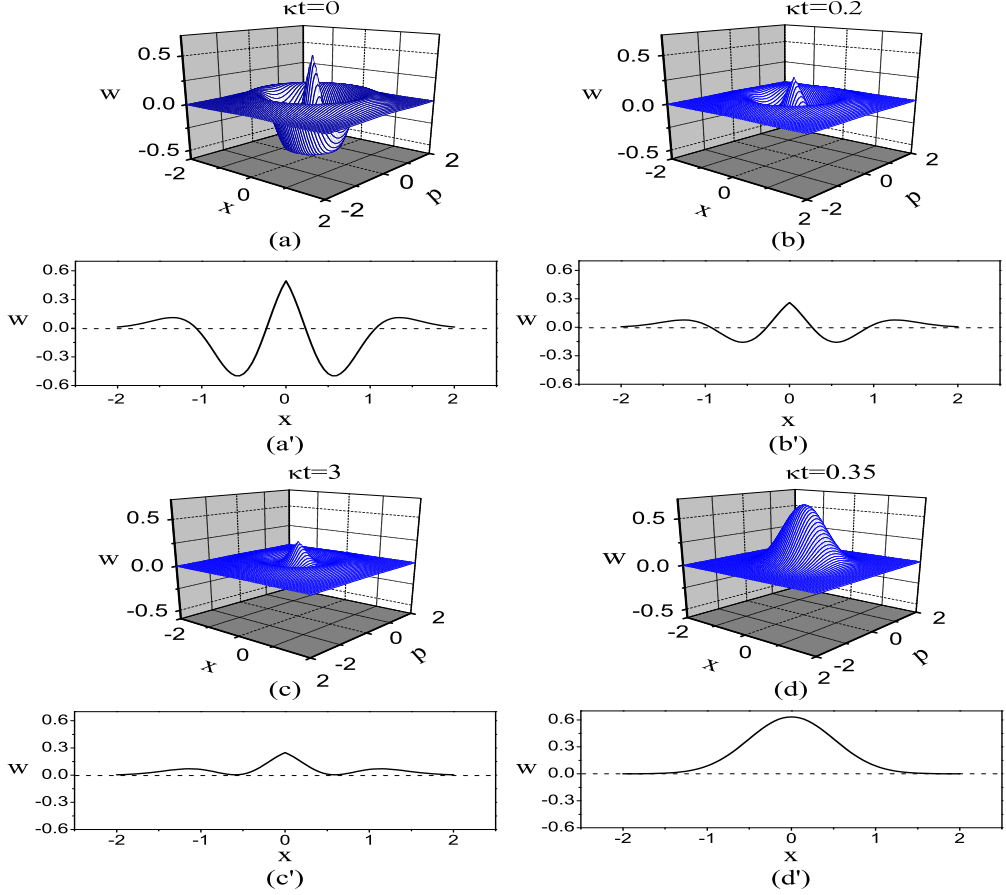}
\end{minipage}
\caption{Wigner functions versus phase space points, $(x,p)$ (upper
line) and $(x,p=0)$ (lower line), of the few-photon superposed state
(3) for different decay times, i.e., $\kappa t=0 (a,a'), 0.2 (b,b'),
0.35 (c,c'), 3 (d,d')$. Here, the parameters in $|\psi(0)\rangle$
are taken as: $|C_{1}|=1/3, |C_{2}|=\sqrt{2}/2, \theta=\varphi=\pi,
\phi=0$.} \label{GWwithP}
\end{figure*}

(i) Initially, the Wigner function shows obviously a negativity,
i.e., at certain phase space points, $W(x,p)<0$. This means that
certain nonclassical effects (such as the anti-bunching of photons)
can be revealed in this initial cavity state.

(ii) With the cavity dissipation, the state of the cavity decays and
the negativity of its time-dependent Wigner function vanishes
gradually. This implies that the nonclassical properties possessed
initially would be weakened with the dissipation of the cavity.

(iii) After certain times, e.g., $\kappa t\geq0.35$ in Fig.~1(c),
the values of the Wigner functions reveal the expected
non-negativity, i.e., $W(x,p)\geq 0$. In this evolved state the
decayed cavity should be classical and the corresponding Wigner
functions could be explained as the usual probabilistic
distributions.

(iv) After the sufficiently-long dissipative time, the cavity state
will decay to the expectable vacuum state or thermal state with the
mean photon number being zero (i.e., $\bar{n}=0$). The Wigner
function for such a dissipated final state should be a Gaussian
distribution. Indeed, from Eq.~(18), we have
\begin{eqnarray}
W(\alpha,\alpha^{*},\infty)&=&\frac{2}{\pi}e^{(-2|\alpha|^{2})}[|C_{0}|^{2}+|C_{1}|^{2}L_{1}^{0}(0)
+|C_{2}|^{2}L_{2}^{0}(0)]\nonumber\\
&=&\frac{2}{\pi}e^{(-2|\alpha|^{2})}[|C_{0}|^{2}+|C_{1}|^{2}
+|C_{2}|^{2}]\nonumber\\
 &=&\frac{2}{\pi}e^{(-2|\alpha|^{2})}.
\end{eqnarray}

\section{Dissipative-dependent quantum statistical properties of
the few-photons cavity initial state}

Various nonclassical effects, e.g., squeezings on quantum
fluctuations and sub-Poisson photon statistics, in quantum optical
states have attracted considerable and continuing interests[24-26].
Many attentions have been paid to find various non-classical optical
states, while how these non-classical effects change with the decays
of the selected non-classical states is a relatively-new topic.
Recently, Biswas and Agarwal discussed how the Mandel $Q$-factor
decreases with the decays of the photon-subtracted squeezed
states[12]. Their numerical results showed that the $Q$-factor
vanishes at the long dissipative times (i.e., $\kappa
t\rightarrow\infty$) and the initial cavity state will decay to the
vacuum. With the dissipative-dependent Wigner functions obtained in
the previous section, we can investigate how the photonic
anti-bunching effect changes with the decay of the few-photon
superposition state$|\psi(0)$ defined in Eq.~(3).

It is well-known that, if the second-order correlation function
\begin{equation}
g^{(2)}(0)=\frac{\langle\hat{a}^{\dagger2}\hat{a}^{2}\rangle}
{\langle\hat{a}^{\dagger}\hat{a}\rangle^{2}}
\end{equation}
is less than $1$, then the photonic distribution in the state
$|\psi\rangle$ is anti-bunching; otherwise, it is bunching. The
symbol $\langle\hat{O}\rangle$ represents the expectation value of
the operator $\hat{O}$ in a quantum state $\rho$. For the present
case we need to calculate the time-dependent expectation values of
the operators $\hat{a}^{\dagger 2}\hat{a}^2$ and
$\hat{a}^{\dagger}\hat{a}$ for the decaying cavity state with
time-dependent Wigner function $W(\alpha,\alpha^*,t)$.

Formally, for an operator function [16]
\begin{eqnarray}
O(\hat{a},\hat{a}^{\dagger})(t)&=&\sum\limits_{n,m}C_{n,m}\hat{a}^{\dagger
n}(t) \hat{a}^{m}(t),
\end{eqnarray}
we have
\begin{eqnarray}
\langle
O(\hat{a},\hat{a}^{\dagger})\rangle(t)&=&Tr[O(\hat{a},\hat{a}^{\dagger})\rho(t)]=
\int d^{2}\alpha O_{S}(\alpha,\alpha^{*}) W(\alpha,\alpha^{*},t).
\end{eqnarray}
On the other hand, from
\begin{eqnarray}
\langle\hat{a}^{\dagger}\rangle(t)&=&[\frac{\partial}{\partial\lambda}
+\frac{\lambda^{*}}{2}]C(\lambda,\lambda^{*},t)|_{\lambda=\lambda^{*}=0}\nonumber\\
\langle\hat{a}\rangle(t)&=&[-\frac{\partial}{\partial\lambda^{*}}
-\frac{\lambda}{2}]C(\lambda,\lambda^{*},t)|_{\lambda=\lambda^{*}=0},
\end{eqnarray}
we can find that
\begin{eqnarray}
\langle
O(\hat{a},\hat{a}^{\dagger})\rangle(t)&=&\sum\limits_{n,m}C_{n,m}[\frac{\partial}{\partial\lambda}
+\frac{\lambda^{*}}{2}]^{n}[-\frac{\partial}{\partial\lambda^{*}}
-\frac{\lambda}{2}]^{m}C(\lambda,\lambda^{*},t)|_{\lambda=\lambda^{*}=0}\nonumber\\
&=&\int
d^{2}\alpha\sum\limits_{n,m}C_{n,m}[\frac{\partial}{\partial\lambda}
+\frac{\lambda^{*}}{2}]^{n}[-\frac{\partial}{\partial\lambda^{*}}
-\frac{\lambda}{2}]^{m}e^{(-\alpha\lambda^{*}+\alpha^{*}\lambda)}|_{\lambda=\lambda^{*}=0}W(\alpha,\alpha^{*},t)\nonumber\\
&=&\int d^{2}\alpha O_{S}(\alpha,\alpha^{*}) W(\alpha,\alpha^{*},t).
\end{eqnarray}
Comparing (24) and (26), we obtain
\begin{equation}
O_{S}(\alpha,\alpha^{*})=\sum\limits_{n,m}C_{n,m}[\frac{\partial}{\partial\lambda}
+\frac{\lambda^{*}}{2}]^{n}[-\frac{\partial}{\partial\lambda^{*}}
-\frac{\lambda}{2}]^{m}e^{(-\alpha\lambda^{*}+\alpha^{*}\lambda)}|_{\lambda=\lambda^{*}=0}.
\end{equation}
Specifically, if $\hat{O}=\hat{a}^\dagger\hat{a}$, then
\begin{equation}
O_{S}(\alpha,\alpha^{*})|_{\hat{O}=\hat{a}^\dagger\hat{a}}=|\alpha|^{2}-\frac{1}{2},
\end{equation}
and thus
\begin{eqnarray}
\langle\hat{a}^{\dagger}\hat{a}\rangle(t)&=&\int d^{2}\alpha
W(\alpha,\alpha^{*},t)O_{S}(\alpha,\alpha^{*})|_{\hat{O}=\hat{a}^{\dagger
2}\hat{a}^2}\nonumber\\
&=& 4|C_{2}|^{2}e^{(-4\kappa t)}+2|C_{2}|^{2}e^{(-2\kappa
t)}[1-2e^{(-2\kappa t)}]+|C_{1}|^{2}e^{(-2\kappa t)},
\end{eqnarray}
Also, if $\hat{O}=\hat{a}^{\dagger 2}\hat{a}^2$, then
\begin{equation}
O_{S}(\alpha,\alpha^{*})|_{\hat{O}=\hat{a}^{\dagger
2}\hat{a}^2}=\frac{1}{2}-2|\alpha|^{2}+|\alpha|^{4},
\end{equation}
and thus
\begin{eqnarray}
\langle\hat{a}^{\dagger 2}\hat{a}^2\rangle(t)&=&\int d^{2}\alpha
W(\alpha,\alpha^{*},t)O_{S}(\alpha,\alpha^{*})|_{\hat{O}=\hat{a}^{\dagger}\hat{a}}\nonumber\\
&=&2|C_{2}|^{2}e^{(-4\kappa t)}.
\end{eqnarray}
Above, the dissipative-dependent Wigner function shown in Eq.~(18)
was used. Consequently, we have
\begin{eqnarray}
g^{(2)}(0;t)&=&\frac{\langle\hat{a}^{\dagger2}\hat{a}^{2}\rangle(t)}{[\langle
\hat{a}^{\dagger}\hat{a}\rangle(t)]^{2}}\nonumber\\
&=&\frac{2|C_{2}|^{2}e^{(-4\kappa t)}}{\{4|C_{2}|^{2}e^{(-4\kappa t)}+2|C_{2}|^{2}e^{(-2\kappa t)}[1-2e^{(-2\kappa t)}]+|C_{1}|^{2}e^{(-2\kappa t)}\}^{2}}\nonumber\\
&=&\frac{2|C_{2}|^{2}}{[|C_{1}|^{2}+2|C_{2}|^{2}]^{2}}=g^{(2)}(0).
\end{eqnarray}
This indicates that the normally-order correlation function
$g^{(2)}(0;t)$ is cavity-loss-invariant; its value depends only on
the initial cavity state!. This is a surprise argument; imagining
that the photons in the initial cavity state is anti-bunching (i.e.,
$g^{(2)}(0;t)<1$), then such a non-classical feature is kept
unchanged even the state approached finally to the vacuum with
non-negative Wigner function. This argument is verified numerically
by Fig.~2(a), which really shows that the value of $g^{(2)}(0;t)$ is
really unchanged with the decay. It is noted that, at the exact
vacuum $|0\rangle$ the expected value of operator
$\langle\hat{a}^\dagger\hat{a}\rangle$ is zero and thus the
definition of $g^{(2)(0)}$ for this state is bizarre and
insignificant. Therefore, our discussion always works for the
dissipative process approaching to (but not arriving at) the exact
vacuum.

\vspace{0.3cm}

\begin{figure}
\includegraphics[width=10.0cm,height=4.0cm]{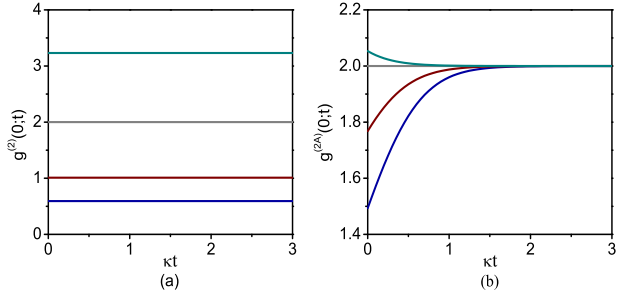}
\caption{(a): Normal-ordered correlation function $g^{(2)}(t)$ is
unchanged with the decay of the few-photons cavity state; (b):
Anti-normally-order correlation function $g^{(2A)}(t)$ versus the
effective decay time of the cavity. Here, the relevant parameters
are taken as: $\theta=\varphi=\pi,\phi=0$, and $|C_{1}|=\sqrt{6}/6,
|C_{2}|=\sqrt{6}/3$ (blue line), $|C_{1}|=2/9,|C_{2}|=2/3$ (red
line), $|C_{1}|=1/3, |C_{2}|=1/3$ (gray line), and $|C_{1}|=1/5,
|C_{2}|=1/3$ (green line), respectively.} \label{GWwithP}
\end{figure}

The dissipative-independence of the normally-correlation function
$g^{(2)}$ can also be proven analytically from the master equation
(7). In fact, at any time $t$ we have
\begin{eqnarray}
\langle\hat{a}^{\dagger}\hat{a}\rangle(t)&=&Tr[\hat{a}^{\dagger}\hat{a}\rho(t)]\nonumber\\
&=&\langle0|\hat{a}^{\dagger}\hat{a}\rho(t)|0\rangle
+\langle1|\hat{a}^{\dagger}\hat{a}\rho(t)|1\rangle
+\langle2|\hat{a}^{\dagger}\hat{a}\rho(t)|2\rangle+...+\langle n|\hat{a}^{\dagger}\hat{a}\rho(t)|n\rangle+...\nonumber\\
&=&0+\langle1|\rho(t)|1\rangle+2\langle2|\rho(t)|2\rangle+...+n\langle n|\rho(t)|n\rangle+...\nonumber\\
&=&\rho_{11}(t)+2\rho_{22}(t)+...+n\rho_{nn}(t)+...\nonumber\\
&=&\sum\limits_{n=0}^\infty n\rho_{n,n}(t),
\end{eqnarray}
and
\begin{eqnarray}
\langle\hat{a}^{\dagger2}\hat{a}^{2}\rangle(t)&=&Tr[\hat{a}^{\dagger}\hat{a}^{\dagger}\hat{a}\hat{a}\rho(t)]
=Tr[\hat{a}^{\dagger}(\hat{a}\hat{a}^{\dagger}-1)\hat{a}\rho(t)]\nonumber\\
&=&[\langle0|\hat{a}^{\dagger}\hat{a}\hat{a}^{\dagger}\hat{a}\rho(t)|0\rangle
+\langle1|\hat{a}^{\dagger}\hat{a}\hat{a}^{\dagger}\hat{a}\rho(t)|1\rangle
+...+\langle n|\hat{a}^{\dagger}\hat{a}\hat{a}^{\dagger}\hat{a}\rho(t)|n\rangle+...]-\langle\hat{a}^{\dagger}\hat{a}\rangle(t)\nonumber\\
&=&[\rho_{11}(t)+2^{2}\rho_{22}(t)+...+n^{2}\rho_{nn}(t)+...]-\langle\hat{a}^{\dagger}\hat{a}\rangle(t)\nonumber\\
&=&\sum\limits_{n=0}^\infty n(n-1)\rho_{n,n}(t),
\end{eqnarray}
and thus
\begin{eqnarray}
g^{(2)}(0;t)&=&\frac{\langle\hat{a}^{\dagger2}\hat{a}^{2}\rangle(t)}{[\langle
\hat{a}^{\dagger}\hat{a}\rangle(t)]^{2}}\nonumber\\
&=&\frac{\langle n^{2}\rangle(t)-\langle n\rangle(t)}{[\langle n\rangle(t)]^{2}}\nonumber\\
&=&\frac{[\rho_{11}(t)+2^{2}\rho_{22}(t)+...+n^{2}\rho_{nn}(t)+...]
-[\rho_{11}(t)+2\rho_{22}(t)+... +n\rho_{nn}(t)+...]}{[\rho_{1}(t)+
2\rho_{22}(t)+...+n\rho_{nn}(t)+...]^{2}}\nonumber\\
&=&\frac{\sum\limits_{n=0}(n^{2}-n)\rho_{nn}(t)}
{[\sum\limits_{n=0}n\rho_{nn}(t)]^{2}}.
\end{eqnarray}
Above, $\rho_{n,n}(t)$ is the diagonal elements of the density
matrix $\rho(t)$ in the Fock space. For the loss cavity initially
prepared in the few-photon superposition state (3), one can easily
see that $\rho_{n,n}=0$, for $n>2$, and the other non-zero diagonal
elements are determined by the following equation (from Eq.~(7)),
\begin{eqnarray}
&&\dot{\rho}_{00}(t)=2\kappa\rho_{11}(t),\nonumber\\
&&\dot{\rho}_{11}(t)=-2\kappa\rho_{11}(t)+4\kappa\rho_{22}(t),\nonumber\\
&&\dot{\rho}_{22}(t)=-4\kappa\rho_{22}(t).
\end{eqnarray}
The solutions to these equations are
\begin{eqnarray}
&&\rho_{11}(t)=[\rho_{11}(0)+2\rho_{22}(0)]e^{-2\kappa t}-2\rho_{22}(0)e^{-4\kappa t}\nonumber\\
&&\rho_{22}(t)=\rho_{22}(0)e^{-4\kappa t}.
\end{eqnarray}
Consequently,
\begin{eqnarray}
g^{(2)}(0;t)&=&\frac{2\rho_{22}(t)}{[\rho_{11}(t)+2\rho_{22}(t)]^{2}}\nonumber\\
&=&\frac{2\rho_{22}(0)e^{-4\kappa t}}{[\rho_{11}(0)e^{-2\kappa t}+2\rho_{22}(0)e^{-2\kappa t}]^{2}}\nonumber\\
&=&\frac{2\rho_{22}(0)}{[\rho_{11(0)}+2\rho_{22}(0)]^{2}}=g^{(2)}(0).
\end{eqnarray}

Suppose that any non-classical effect should vanish due to the
dissipation, the dissipative-independence of the
normally-correlation function implies that such a parameter should
not be a good physical quantity to describe the cavity loss.
Alternatively, the anti-normal ordered correlation function, defined
as
\begin{equation}
g^{(2A)}(0)=\frac{\langle\hat{a}^{2}\hat{a}^{\dagger2}\rangle}{\langle
\hat{a}\hat{a}^{\dagger}\rangle^{2}}
=\frac{\langle\hat{a}^{\dagger2}\hat{a}^{2}\rangle
+4\langle\hat{a}^{\dagger}\hat{a}\rangle+2}{[\langle
\hat{a}^{\dagger}\hat{a}\rangle+1]^{2}},
\end{equation}
could be utilized to describe the dissipative process of the
few-photon cavity. Indeed, with Eqs.~(29) and (31) we have
\begin{eqnarray}
g^{(2A)}(0;t)&=&\frac{\langle\hat{a}^{\dagger2}\hat{a}^{2}\rangle(t)+4\langle\hat{a}^{\dagger}\hat{a}\rangle(t)+2}{[\langle
\hat{a}^{\dagger}\hat{a}\rangle(t)+1]^{2}}\nonumber\\
&=&\frac{2|C_{2}|^{2}e^{(-4\kappa t)}+ 4\{4|C_{2}|^{2}e^{(-4\kappa
t)} +2|C_{2}|^{2}e^{(-2\kappa t)}[1-2e^{(-2\kappa t)}]
+|C_{1}|^{2}e^{(-2\kappa t)}\}+2}{\{4|C_{2}|^{2}e^{(-4\kappa t)}
+2|C_{2}|^{2}e^{(-2\kappa t)}[1-2e^{(-2\kappa t)}]
+|C_{1}|^{2}e^{(-2\kappa
t)}+1\}^{2}}\nonumber\\
&=&\frac{4|C_{1}|^{2}e^{(-2\kappa t)}+8|C_{2}|^{2}e^{(-2\kappa
t)}+2|C_{2}|^{2}e^{(-4\kappa t)}+2}{[|C_{1}|^{2}e^{(-2\kappa
t)}+2|C_{2}|^{2}e^{(-2\kappa t)}+1]^{2}},
\end{eqnarray}
which is not an invariant during the cavity dissipation. One can see
also from Fig.~2(b) that, the value of the anti-normal correlation
function changes with the cavity loss. After a sufficiently-long
decay time the value of $g^{(2A)}(0;t)$ should limit to $2$,
whatever its initial value is less than $2$ or not.
Certainly, such a dissipative-dependent behavior of the
$g^{(2A)}(0;t)$-parameter can also be exactly verified by using the
analytic solutions, i.e., Eq.~(37). In fact, we can see that
\begin{eqnarray}
g^{(2A)}(0;t)&=&\frac{\langle\hat{a}^{\dagger2}\hat{a}^{2}\rangle(t)+4\langle\hat{a}^{\dagger}\hat{a}\rangle(t)+2}{[\langle
\hat{a}^{\dagger}\hat{a}\rangle(t)+1]^{2}}\nonumber\\
&=&\frac{4\rho_{11}(0)e^{-2\kappa t}+8\rho_{22}(0)e^{-2\kappa
t}+2\rho_{22}(0)e^{-4\kappa t}+2} {[\rho_{11}(0)e^{-2\kappa
t}+2\rho_{22}(0)e^{-2\kappa t}+1]^{2}}.
\end{eqnarray}
It is consistent with the Eq.~(41), as if $t\rightarrow\infty$,
Eq.(42) can be shown
\begin{equation}
g^{(2A)}(0; t\rightarrow\infty)=2.
\end{equation}

\section{Possible experimental verification:
the preparation of few-photon superposed states and measurement of
its Wigner function}

We now discuss how to test the above arguments with a typical cavity
QED system, i.e., highly excited Rydberg atoms in a high-Q microwave
cavity [28]. An ideal setup is schematized in Fig.~4, wherein an
atom is emitted from the source O and then flies across sequentially
a quantized cavity, a classical microwave field, and finally is
detected in the detector $I$.
When the atom passes through the quantized cavity, the usual
Jaynes-Cummings model with the Hamiltonian ($\hbar=1$)
\begin{equation}
H=\omega_{a}S_{z}+\omega_{c}\hat{a}^{\dagger}\hat{a}+g(\hat{a}S_{+}+\hat{a}^{+}S_{-}),
\end{equation}
works. Here, $\omega_{a}$, $\omega_{c}$ are the atomic transition
frequency and the cavity field frequency, respectively. $S_{z}$,
$S_{\pm}$ are the atomic operators, such that $[S_{+},S_{-}]=2S_{z},
[S_{z},S_{\pm}]=\pm S_{\pm}$. $\hat{a}$ and $\hat{a}^{\dagger}$ are
the annihilation and creation operators of the cavity field,
respectively. And, $g$ is the atom-field coupling strength.
\begin{figure}
\centering
\includegraphics[width=3in,height=1.5in]{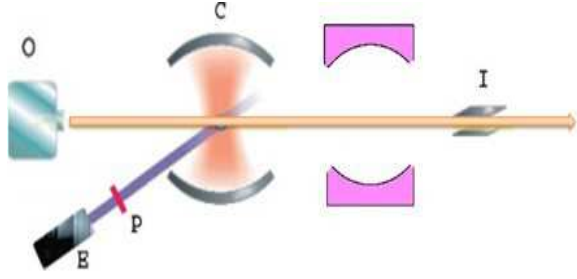}\\
\caption{An experimental setup for preparing the superposition
states of $|0\rangle, |1\rangle$ and $|2\rangle$. Here, an atom is
emitted from the source O, then it flies sequentially across the J-C
cavity, the classical microwave field, and at last is detected in
the detector $I$.}\label{winger2}
\end{figure}

Initially, the atom is in the ground state $|e_{1}\rangle$ and the
cavity mode in the vacuum state, i.e., the wave function of the
atom-cavity system is $|\psi(0)\rangle=|0, e_{1}\rangle$. Next, the
atom is injected into the cavity and the state of the atom-cavity
system evolves to
\begin{equation}
|\psi(t)\rangle_{1}=\cos(gt_{1})|0,e_{1}\rangle-i\sin(gt_{1})|1,g_{1}\rangle,
\end{equation}
after the passage time $t_1$. Then, we let the atom continuously
across a classical microwave field for evolving the atomic states
as: $|e_{1}\rangle\longrightarrow\cos(\theta_{1}/2)|e_{1}\rangle\
+ie^{-i\varphi_{1}}\sin(\theta_{1}/2)|g_{1}\rangle$ and
$|g_{1}\rangle\longrightarrow\cos(\theta_{1}/2)|g_{1}\rangle\
+ie^{i\varphi_{1}}\sin(\theta_{1}/2)|e_{1}\rangle$. Here, the values
of $\theta_{1}$ and $\varphi_{1}$ are adjustable. Therefore, before
arriving at the atomic detector I, the state of the atom-cavity
system reads
\begin{eqnarray}
|\psi(t)\rangle_{1}&=&[\cos(g t_{1})\cos(\frac{\theta_{1}}{2})|0\rangle
+e^{i\varphi_{1}}\sin(g t_{1})\sin(\frac{\theta_{1}}{2})|1\rangle]|e_{1}\rangle\nonumber\\
&+&[ie^{-i\varphi_{1}}\cos(gt_{1})\sin(\frac{\theta_{1}}{2})|0\rangle
-i\sin(gt_{1})\cos(\frac{\theta_{1}}{2})|1\rangle]|g\rangle.
\end{eqnarray}
In order to generate the desirable superposition of the states
$|0\rangle, |1\rangle$ and $|2\rangle$, we must let another atom (as
the same of the former one) pass sequentially across the cavity and
the microwave field. Finally, the state of the whole system
including two atoms and a cavity mode can be expressed as:
\begin{eqnarray}
|\psi(t)\rangle_{2}&=&|0\rangle\{\cos(gt_{1})\cos(gt_{2})\cos\frac{\theta_{1}}{2}\cos\frac{\theta_{2}}{2}|e_{1}\rangle|e_{2}\rangle     \nonumber\\
&+&ie^{-i\varphi_{2}}\cos(gt_{1})\cos(gt_{2})\cos\frac{\theta_{1}}{2}\sin\frac{\theta_{2}}{2}]|e_{1}\rangle|g_{2}\rangle\}     \nonumber\\
&+&|1\rangle\{\sin(gt_{1})\cos(gt_{2})\sin\frac{\theta_{1}}{2}\cos\frac{\theta_{2}}{2}e^{i\varphi_{1}}|e_{1}\rangle|e_{2}\rangle\nonumber\\
&+&ie^{-i\varphi_{1}}e^{i\varphi_{2}}\sin(gt_{1})\cos(gt_{2})\sin\frac{\theta_{1}}{2}\sin\frac{\theta_{2}}{2}|e_{1}\rangle|g_{2}\rangle\nonumber\\
&-&i\cos(gt_{1})\sin(gt_{2})\cos\frac{\theta_{1}}{2}\cos\frac{\theta_{2}}{2}|e_{1}\rangle|g_{2}\rangle  \nonumber\\
&+&e^{i\varphi_{2}}\cos(gt_{1})\sin(gt_{2})\cos\frac{\theta_{1}}{2}\sin\frac{\theta_{2}}{2}|e_{1}\rangle|e_{2}\rangle\}   \nonumber\\
&+&|2\rangle\{e^{i\varphi_{1}}e^{i\varphi_{2}}\sin(gt_{1})\sin(gt_{2})\sin\frac{\theta_{1}}{2}\sin\frac{\theta_{2}}{2}|e_{1}\rangle|e_{2}\rangle     \nonumber\\
&-&ie^{i\varphi_{1}}\sin(gt_{1})\sin(gt_{2})\sin\frac{\theta_{1}}{2}\cos\frac{\theta_{2}}{2}|e_{1}\rangle|g_{2}\rangle\},
\end{eqnarray}
As a consequence, the desirable few-photon superposed state can be
generated by the state-selective measurements on the atoms. For
example, if the atoms are detected at the state
$|e_1\rangle|e_2\rangle$, then the cavity mode collapses into
\begin{eqnarray}
|\psi(t)\rangle_2&=&
\frac{1}{\sqrt{N}}\{|0\rangle[\cos(gt_{1})\cos(gt_{2})\cos\frac{\theta_{1}}{2}\cos\frac{\theta_{2}}{2}]\nonumber\\
&+&|1\rangle[e^{i\varphi_{1}}\sin(gt_{1})\cos(gt_{2})\sin\frac{\theta_{1}}{2}\cos\frac{\theta_{2}}{2}
+e^{i\varphi_{2}}\cos(gt_{1})\sin(gt_{2})\cos\frac{\theta_{1}}{2}\sin\frac{\theta_{2}}{2}]\nonumber\\
&+&|2\rangle[e^{i\varphi_{1}}e^{i\varphi_{2}}\sin(gt_{1})\sin(gt_{2})\sin\frac{\theta_{1}}{2}\sin\frac{\theta_{2}}{2}]\}|e_{1}\rangle|e_{2}\rangle,
\end{eqnarray}
with
\begin{eqnarray}
N&=&[\cos gt_{1}\cos gt_{2}\cos\frac{\theta_{1}}{2}\cos\frac{\theta_{2}}{2}]^{2}+[\sin gt_{1}\sin gt_{2}\sin\frac{\theta_{1}}{2}\sin\frac{\theta_{2}}{2}]^{2}\nonumber\\
&+&[\sin gt_{1}\cos
gt_{2}\sin\frac{\theta_{1}}{2}\cos\frac{\theta_{2}}{2}]^{2}+[\cos
gt_{1}\sin
gt_{2}\cos\frac{\theta_{1}}{2}\sin\frac{\theta_{2}}{2}]^{2}\nonumber\\
&+&\frac{1}{8}\cos(\varphi_{1}-\varphi_{2})\sin 2gt_{1}\sin
2gt_{2}\sin\theta_{1}\sin\theta_{2}
\end{eqnarray}
being the normalized coefficient. If the relevant parameters are set
properly as: $\varphi_{1}=\pi, \varphi_{2}=0,
gt_{1}=gt_{2}=\theta_{2}/2=\pi/4, \theta_{1}/2=7\pi/4$, then a
typical few-photon state discussed above
\begin{eqnarray}
|\psi(t)\rangle_2=\frac{\sqrt{6}}{6}|0\rangle
+\frac{\sqrt{6}}{3}|1\rangle+\frac{\sqrt{6}}{6}|2\rangle
\end{eqnarray}
can be obtained.

The method to measure the Wigner function for a given cavity state
is relative standard. Here, we recommend the approach proposed by
Lutterbach and Davidovich [28] by directly detecting the the
negativity of Wigner function via the atomic Ramsey
interferometries. In fact, at a phase space point $\alpha$, Wigner
function for the cavity state with the density matrix $\rho$ can be
simply expressed by [29]
\begin{equation}
W(\alpha)=2Tr[D(-\alpha)\rho D(\alpha)P]=2\langle P\rangle.
\end{equation}
Here, $P=\exp(i\pi\hat{a}^{+}\hat{a})$ and
$D(\alpha)=\exp(\alpha\hat{a}^{+}-\alpha^{*}\hat{a})$.
Furthermore, the quantity $\langle P\rangle$ can be determined by
measuring the probability $P_{e}$ (or $P_{g}$) of the atom is
detected at its excited state $|e\rangle$ (or $|g\rangle$), i.e.,
\begin{eqnarray}
P_{e}(\phi,\alpha)=\frac{1}{2}[1+\langle P\rangle\cos\phi].
\end{eqnarray}
Therefore, the Wigner function is determined by
\begin{eqnarray}
W(\alpha)&=&2[P_{e}(0,\alpha)-P_{e}(\pi,\alpha)].
\end{eqnarray}
Consequently, if we have
\begin{equation}
P_{e}(0,\alpha)<P_{e}(\pi,\alpha),
\end{equation}
then the Wigner function attains a negative value. With these
preparations and measurements, the dissipative dynamics presented
above could be tested experimentally.

\section{Discussions and Conclusions}

With the few-photon superposed state, in this paper we have
investigated the dissipative dynamics of the quantized mode without
any thermal photon. By numerical method, we discuss how the Wiginer
function of the cavity state changes with the dissipation of the
cavity. Our results show clearly that the initial negativity of the
Wigner function vanishes with the cavity dissipation.
With the dissipative-dependent Wigner function, we further discuss
how a typical quantum statistical property, the second-order
correlation function $g^{(2)}(0)$, changes with the dissipation of
the cavity. It is surprised that such a quantity is an invariant
during the dissipation of the cavity. This argument was also
verified by analytical method directly solving the master equation
of the dissipative cavity. This implies that the $g^{(2)}(0)$ should
not be a good physical quantity to describe the dissipative dynamics
of the cavity, at least for the few-photon state.

The discussion in the present work is limited to the photons in
optical cavity, and thus the mean thermal photons at room
temperature can be really negligible. This implies that the final
state of the dissipative optical cavity is exactly vacuum, at which
the standard definition of the second-order correlation function is
bizarre and insignificant. The generalization to the dissipative
cavity with non-zero thermal photons is in progress.

Given the few-photon superposed state of the cavity is not difficult
to be prepared and its relevant Wigner function can also be easily
measured in the usual cavity QED system, we believe that our
arguments are testable with the current experimental technique.

\section*{Acknowledgments}
One of us (Wen) thanks Dr. W. Z. Jia for useful discussions. This
work was supported in part by the National Science Foundation grant
No. 10874142, 90921010, 11174373 and the National Fundamental
Research Program of China through Grant No. 2010CB923104. J. Cheng
thanks the supports from the National Basic Research Program of
China under Grant No. 2012CB921904, the National Natural Science
Foundation of China (11174084, 10934011), the Fundamental Research
Funds for Central University (SCUT), and the State Key Laboratory of
Precision Spectroscopy.

\end{document}